\documentclass{elsart}

\usepackage{amssymb}
\usepackage{graphicx}
\usepackage{dcolumn}
\usepackage{bm}
\usepackage{color}

\usepackage{psfrag}

\begin{document}

\begin{frontmatter}



\title{Accurate estimator of correlations between asynchronous signals}


\author[isi,bme]{Bence T\'oth}
\author[bme]{J\'anos Kert\'esz}

\address[isi]{ISI Foundation - Viale S. Severo 65, 10133 Torino, Italy}
\address[bme]{Institute of Physics, Budapest University of Technology and
Economics - Budafoki \'ut. 8.  H-1111 Budapest, Hungary}

\begin{abstract}
The estimation of the correlation between time series is often
hampered by the asynchronicity of the signals. Cumulating data within
a time window suppresses this source of noise but weakens the
statistics.  We present a method to estimate correlations without
applying long time windows.  We decompose the correlations of data
cumulated over a long window using decay of lagged correlations as
calculated from short window data.  This increases the accuracy of the
estimated correlation significantly and decreases the necessary
efforts of calculations both in real and computer experiments.
\end{abstract}

\end{frontmatter}



\section{\label{sec:intro} Introduction}
Correlations between time series are fundamental for understanding and
interpreting stochastic processes. Very often the time between the
signals of a series is distributed in an uneven fashion, causing
asynchronicity of the compared series.

Correlations between asynchronous signals can be of great
importance in several areas. An important example is the
case of neutron activation analysis \cite{hevesy1936}. These experiments are
used for non-destructive testing of materials in order to determine the
concentration of their constituents. In the analysis the specimen is bombarded
by neutrons coming from a source, causing the elements to form radioactive
isotopes, and from the spectra of the emissions of the
radioactive sample, the concentration of the elements can be
determined. Since the radiation appears together for all kinds of
atoms, different elements are going to radiate in a correlated but
asynchronous way \cite{hevesy1936,soete1972}.

Another example can be taken from materials science or seismology. Mechanical
failures can be tested by their wave radiation and it is crucial to know if the
signals measured by different sensors are correlated \cite{McConnell1995}.
Obviously, the analysis involves the handling of asynchronous signals.

Correlations of returns of different companies are fundamental input data for
portfolio optimisation. As the transactions are asynchronous, the correlations
measured on short time scale are significantly reduced, which is called the Epps
effect \cite{epps1979}.

From theoretical point of view continuous time random walks \cite{montroll1965}
can be mentioned, that can be used to describe a broad range of processes from
transport in disordered solids \cite{scher1973a} to finance
\cite{scalas2006}. Correlated continuous random walks produce asynchronous time
series.

When computing the Pearson correlation coefficient of two stationary signals we
often have to face an important problem: The correlation measure is designed to
determine the grade of co-movements of synchronous observations, while the
signals are asynchronous.  The usual way to handle this problem is to
cumulate data over a time window $\Delta t$ and look for the correlations
between these binned data. In order to approach the asymptotic, proper value of
the correlation coefficient, $\Delta t$ should be much larger than the scale of
asynchronicity. However, this leads to the reduction of the
statistics, consequently it makes the estimates inaccurate.
On the other hand, for short $\Delta t$ the noise due to
asynchronicity may reduce the measured correlations significantly.

It has been suggested to use measures of correlation other
than the Pearson coefficient to overcome the problem of
asynchronicity. Ref. \cite{malliavin2002} presents a method of
measuring covariance based on Fourier series analysis of data. This
method has been applied by Refs. \cite{reno2003,precup2004}
in the study of financial correlations.
While the Fourier method is indeed somewhat less sensitive to
asynchronicity, the problem cannot be eliminated by its use.
Refs. \cite{hayashi2005} propose a new estimator of the
covariance of two diffusion processes that are observed only at
discrete times in a non-synchronous manner. Their estimator uses all
available data and does not require synchronization of observations,
however in the presence of noise it becomes inconsistent and its
variance diverges. A good comparison of several covariance estimators
can be found in Refs \cite{voev2006,palandri2006}.

In this paper we describe an estimator of the Pearson correlation coefficient
for correlated, asynchronous pairs of data based on an appropriate decomposition
of the expression for the correlation coefficient for large $\Delta t$ by using
the value of the coefficient for small time window and decay of lagged
correlations. The latter can be calculated from the good statistics high
resolution data. We applied the decomposition scheme already to explain
\cite{toth2007b} the Epps effect \cite{epps1979}. Here we show how it can be
used to accurately estimate the asymptotic correlation coefficients.

The paper is organized as follows: In Section \ref{sec:decomp} we
review the decomposition of correlations. Section \ref{sec:method}
contains the description of our method together with some
demonstrative results.
In Section \ref{sec:data} we apply the method to real data.
We end the paper with a short summary in
Section \ref{sec:summary}.

\section{\label{sec:decomp} Decomposition of correlations}

Let us consider a system, where discrete stationary signals arrive
from two different sources (A and B) at different time instances
resulting in two correlated time series. We count the hits arriving,
and denote their cumulative number measured from a reference time
$t=0$ at time $t$ by $c^{A}(t)$ and $c^{B}(t)$ respectively.

We are interested in the correlation between the number of hits
arriving to our sensor in a certain time window, thus the change in
$c^{A}$ and $c^{B}$. This will be denoted by

\begin{equation}
\label{eq:ret}
r^{A}_{\Delta t}(t)=c^{A}(t)-c^{A}(t-\Delta t).
\end{equation}

The general Pearson correlation measure with time lag $\tau$ is
defined by

\begin{eqnarray}
\label{eq:C}
C_{\Delta t}^{A/B}(\tau)=\frac{\left\langle \Delta r_{\Delta t}^{A}(t)\Delta
 r_{\Delta
t}^{B}(t+\tau)\right\rangle}{\sigma^{A}\sigma^{B}},
\end{eqnarray}
where $\Delta r_{\Delta t}(t)$ is the deviation of $r_{\Delta t}(t)$
from its mean,

\begin{eqnarray}\label{eq:time_ave}
\left\langle \Delta r_{\Delta t}^{A}(t)\Delta r_{\Delta
t}^{B}(t+\tau)\right\rangle= \nonumber \\
=\frac{1}{N}\sum_{i=1}^{N} \Delta r_{\Delta t}^{A}(i\Delta t)\Delta r_{\Delta
t}^{B}(i\Delta t +\tau),
\end{eqnarray}
with $N=[(T-\tau)/\Delta t]$ and $\sigma^{A}$ ($\sigma^{B}$) is the
standard deviation of $r_{\Delta t}^{A}$ ($r_{\Delta t}^{B}$).

We use $\langle\dots\rangle$ for denoting time average. The equal-time
 correlation
coefficient is naturally: \(\rho_{\Delta t}^{A/B}\equiv C_{\Delta
t}^{A/B}(\tau=0)\).

As we can see, already by the definition, the length of the time
window, i.e. the value of $\Delta t$ plays a major role in measuring the
correlation.

Assuming $\Delta t = n\Delta t_0$ with $n$ being a positive integer,
we can deduce the following relationship between correlations on the
two different time scales

\begin{eqnarray}
\label{eq:data_formula2}
\rho_{\Delta t}^{A/B}=\Bigg(\sum_{x=-n+1}^{n-1}\left(n-|x|\right)\left\langle
 r_{\Delta t_{0}}^{A}(t)r_{\Delta t_{0}}^{B}(t+x\Delta
 t_{0})\right\rangle-n^2\left\langle r^{A}_{\Delta
 t_{0}}(t)\right\rangle\left\langle r^{B}_{\Delta
 t_{0}}(t)\right\rangle\Bigg)\times
    \nonumber \\
\Bigg(\sum_{x=-n+1}^{n-1}\left(n-|x|\right)\left\langle r_{\Delta
 t_{0}}^{A}(t)r_{\Delta t_{0}}^{A}(t+x\Delta t_{0})\right\rangle-n^2\left\langle
 r^{A}_{\Delta
      t_{0}}(t)\right\rangle^{2}\Bigg)^{-1/2}\times \nonumber \\
\Bigg(\sum_{x=-n+1}^{n-1}\left(n-|x|\right)\left\langle r_{\Delta
 t_{0}}^{B}(t)r_{\Delta t_{0}}^{B}(t+x\Delta t_{0})\right\rangle-n^2\left\langle
 r^{B}_{\Delta t_{0}}(t)\right\rangle^{2}\Bigg)^{-1/2}.
\end{eqnarray}

Details on the deduction of Equation \ref{eq:data_formula2} can
be found in the Appendix. It is plausible to set $\Delta t_{0}$ as the
shortest meaningful time scale in the system, that has to be chosen
with the actual problem in mind.

This decomposition can be used to accurately estimate correlations by
using high resolution, i.e., good statistics data.

\section{\label{sec:method} Method}
Expression (\ref{eq:data_formula2}) enables to calculate the
correlation coefficient for any sampling time scale, $\Delta t$, by
knowing the coefficient on a shorter sampling time scale, $\Delta
t_0$, and the decay of lagged correlations on the same shorter
sampling time scale (given that $\Delta t$ is multiple of $\Delta
t_0$).

The method we would like to propose relies on this decomposition of
correlations.

We suggest the following procedure. Data should be binned with a small
$\Delta t_0$ such that a good statistics is achieved, irrespective of
the fact that noise due to asynchronicity may be considerable. Then
the correlation functions and the decay of lagged correlations should
be calculated using these data (of course the calculated correlation
can be expected to be too small). Plugging in these quantities into
Equation (\ref{eq:data_formula2}) with a large enough $\Delta t$ we
obtain a good estimate for the proper correlation. Using different
values of $\Delta t$ an extrapolation to the proper, asymptotic
correlation coefficient is possible.

We demonstrate the method in more details in this section. We use
correlated random walks and show that in case of directly measuring
the correlations in large $\Delta t$ time windows we need very long
time series in order to have a good estimate for correlations.  When
using the decomposition method, we use high resolution data and thus
can achieve an estimate for the correlations with the same accuracy
from a dataset of much shorter time span.

\subsection{\label{sec:demo} Demonstration}
\begin{figure}
\begin{center}
\psfrag{W(t)}[][][2.5][0]{W(t)}
\psfrag{cA(t)}[][][2.5][0]{$c^{A}(t)$}
\psfrag{cB(t)}[][][2.5][0]{$c^{B}(t)$}
\includegraphics[angle=-90,width=0.750\textwidth]{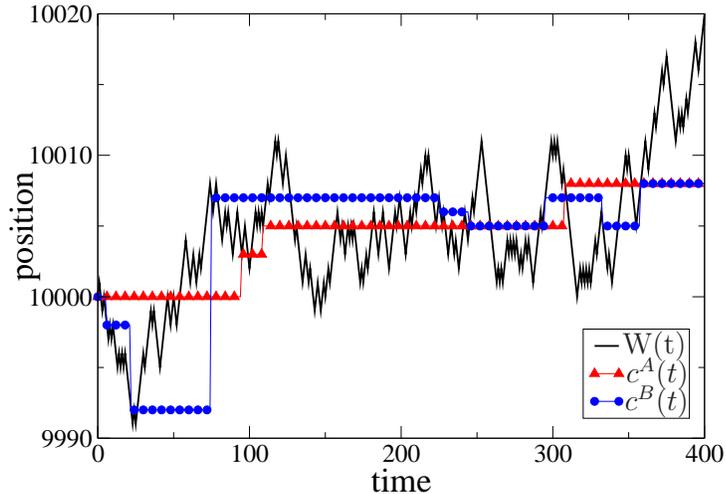}

\caption{\label{fig:model}(Color online) Illustration of the asynchronous
 sampling,
as introduced in \cite{toth2007b}. The original random walk is shown with
lines (black), the two sampled series with triangles and line (red)
and dots and line (blue).}
\end{center}
\end{figure}

We construct correlated asynchronous time series in the following way:
As a first step we generate a core random walk with unit steps up or
down in each second:

\begin{eqnarray}
\label{eq:model_def1}
W(t)= W(t-1)+\varepsilon(t),
\end{eqnarray}

where $\varepsilon(t)$ is $\pm 1$ with equal probability. Second we sample
the random walk, $W(t)$, twice independently with sampling time intervals drawn
from some distribution. This way we simulate asynchronicity by non-simultaneously
sampling our generated data. A snapshot of the random walks can be seen in Figure
\ref{fig:model}.

Below we present results first for exponentially distributed sampling intervals
between steps and second for the sampling intervals between steps being
drawn from a Weibull distribution.  We study the correlation between
the changes in the position of the two random walkers.

\subsubsection{\label{sec:exp} Exponential sampling intervals}
\begin{figure}
\begin{center}
\includegraphics[angle=0,width=0.75\textwidth]{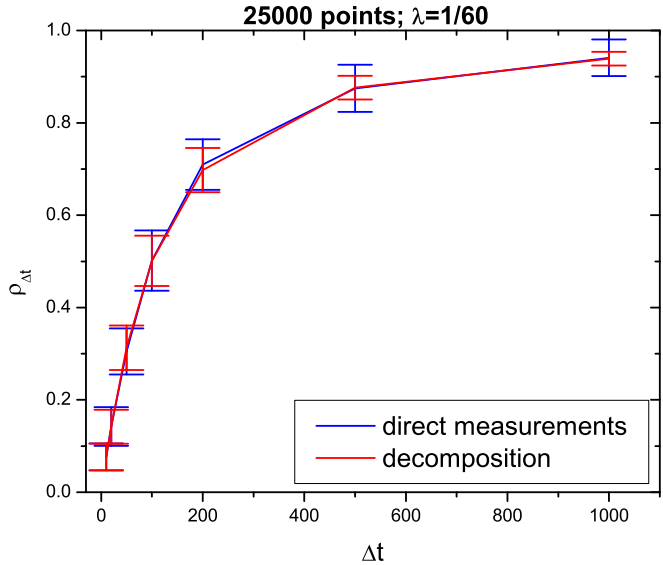}
\caption{\label{fig:exp} (Color online) Comparison of the directly measured
correlation coefficients and the coefficients determined through the
decomposition method in case of exponentially distributed sampling intervals.
In blue we show the average and the standard deviation
of the direct measurements computed over the ensemble of 50 time series
pairs. In red we show the average and the standard deviation
obtained by the decomposition method, computed over the ensemble of 50 time
 series pairs.}
\end{center}
\end{figure}

We generate 50 pairs of time series, as described above, with sampling intervals

between consecutive changes from the distribution:

\begin{eqnarray}
\label{eq:exp}
\mathbb{P}(y)=\Bigg\{\begin{array}{ll}
\lambda e^{-\lambda y} & \textrm{if } y \ge 0 \\
0 & y<0
\end{array}
\end{eqnarray}

\noindent with parameter $\lambda=1/60$. Each time series has a length of 25000
time steps. Naturally, since the time series are finite, the
correlation coefficients that we measure will have errors.
In Figure \ref{fig:exp} we show the results for the correlation
coefficients on different time scales, where the shortest time scale
used was $\Delta t_{0}=10$.

In blue we show the average of the direct measurements, taken
over 50 points. The errorbars show the standard deviation of the points.
In red we show the average result obtained by the decomposition
 method
using Equation \ref{eq:data_formula2},
taken over 50 points, with their standard deviation as errorbars.
As we can see there is a significant difference between the errors of the two
 measurements,
while their means are very near to each other. In general, the error of
 measurements
goes as $\sigma/\sqrt{N}$, where
$\sigma$ is the standard deviation of the distribution of results and
$N$ is the number of data points. The ratio of the standard deviations at $\Delta t=1000$
in Figure \ref{fig:exp} is close to 3.
This means that in order to obtain the same precision from direct
measurements as from the decomposition method, we need roughly one order
of magnitude more data points.

Generally, one is interested in the asymptotic value of the
correlations, i.e. in the limit of $\rho_{\Delta t \to \infty}$. As we
can see, even for the numerically generated data, determining the
correlation for the scale of $\Delta t=1000$ we still get a
correlation different from the underlying asymptotic correlation, that
is 1. A possible way to determine the asymptotic correlation value is
using some extrapolation method.

\begin{figure}
\begin{center}
\includegraphics[angle=0,width=0.75\textwidth]{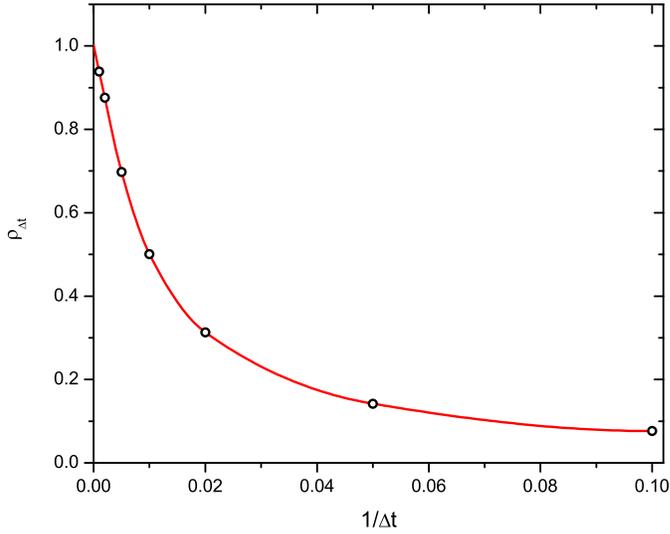}
\caption{\label{fig:exp_poly} (Color online) The correlation as a function
 of
$1/\Delta t$ for the exponentially distributed sampling intervals. The circles
show the correlations determined using the decomposition method, and
the (red) curve is the piecewise Cubic Hermite Interpolation of the
correlation values. The extrapolation gives an asymptotic correlation
value of 1.002, very close to the actual underlying value, that is 1.}
\end{center}
\end{figure}

We demonstrate that applying a simple extrapolation method for
the generated data we can determine the exact underlying correlation
with very good accuracy. Since we are interested in the
$\Delta t \to \infty$ value, we use the plot of $\rho_{\Delta t}$ as a
function of $1/\Delta t$ for the extrapolation. Figure
\ref{fig:exp_poly} shows the correlation points and the curve
determined by piecewise Cubic Hermite Interpolation method. The
extrapolated curve intercepts the y-axis at the value of 1.002, which
is very close to the actual asymptotic correlation value, with an
error around $0.2\%$.

We studied the effect of the errors of the two methods on the accuracy of
the extrapolated value of the correlation coefficient. Applying a  piecewise
Cubic Hermite Interpolation to the endpoints of the error bars, we find that
the extrapolated value of the asymptotic correlations using direct measurements
falls between 0.979 and 1.035, while the value for the decomposition method
falls between 0.998 and 1.005, indicating a factor of 8 improvement in the precision.


\subsubsection{\label{sec:weibull} Weibull sampling intervals}
\begin{figure}
\begin{center}
\includegraphics[angle=0,width=0.75\textwidth]{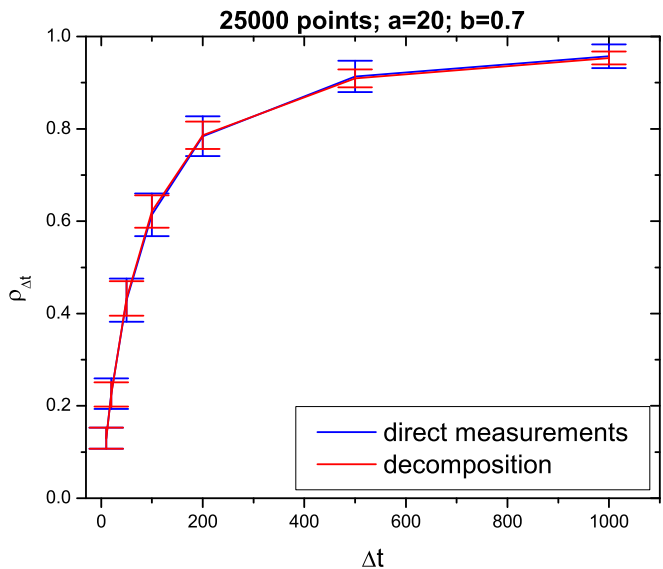}
\caption{\label{fig:weibull} (Color online) Comparison of the directly measured
correlation coefficients and the coefficients determined through the
decomposition method in case of Weibull distributed sampling intervals.
In blue we show the average and the standard deviation
of the direct measurements computed over the ensemble of 50 time series
pairs. In red we show the average and the standard deviation
obtained by the decomposition method, computed over the ensemble of 50 time
 series pairs.}
\end{center}
\end{figure}

In order to demonstrate the power of the method for a non Poisson
process we generate 50 pairs of time series, as described above, with
sampling intervals between consecutive changes generated from a Weibull
distribution:

\begin{eqnarray}
\label{eq:weibull}
\mathbb{P}(y)=\Bigg\{\begin{array}{ll}
\frac{b}{a}(\frac{y}{a})^{b-1}e^{-(y/a)^{b}}& \textrm{if } y \ge 0 \\
0 & y<0
\end{array}
\end{eqnarray}

with parameters $a=20$ and $b=0.7$. Again each time series has a
length of 25000 time steps and the directly measured correlation
values have large variance. In Figure \ref{fig:weibull} we show the
results for the correlation coefficients on different time scales,
again with $\Delta t_{0}=10$.

\begin{figure}
\begin{center}
\includegraphics[angle=0,width=0.75\textwidth]{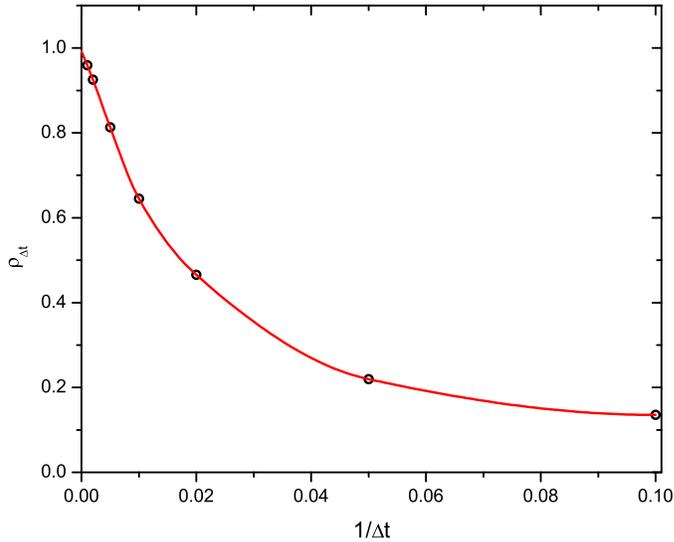}
\caption{\label{fig:weibull_poly} (Color online) The correlation as a function
 of
$1/\Delta t$ for the Weibull distributed sampling intervals. The circles
show the correlations determined using the decomposition method, and
the (red) curve is the piecewise Cubic Hermite Interpolation of the
correlation values. The extrapolation gives an asymptotic correlation
value of 0.992, very close to the actual underlying value, that is 1.}
\end{center}
\end{figure}

The Figure shows that we get
a better estimate of the means from the decomposition formula than
from direct measurements. In this case the ratio of the standard deviations
at $\Delta t=1000$ is close to 2. The same improvement is obtained for the
precision of the extrapolated asymptotic correlation coefficients. The
extrapolation of the asymptotic coefficient can be seen in Figure \ref{fig:weibull_poly}.

The two above examples on generated time series show that using our
method and estimating the correlation coefficient between asynchronous
signals from the high frequency data leads to much smaller variation
of the results than in case of direct measurement of correlations on
lower frequency data.

\subsection{\label{sec:decay} Decay functions}

As we can see from Equation (\ref{eq:data_formula2}) a subtle point
of the correlation estimation is the measurement of the decay of
lagged correlations on the short ($\Delta t_{0}$) time scale, that we
will call decay functions. There are three decay functions we have to
measure: $\left\langle r_{\Delta t_{0}}^{A}(t)r_{\Delta
t_{0}}^{A}(t+x\Delta t_{0})\right\rangle$, $\left\langle r_{\Delta
t_{0}}^{B}(t)r_{\Delta t_{0}}^{B}(t+x\Delta t_{0})\right\rangle$ and
$\left\langle r_{\Delta t_{0}}^{A}(t)r_{\Delta t_{0}}^{B}(t+x\Delta
t_{0})\right\rangle$. To have good estimation of the asymptotic
correlation value, one has to have precise measurements of these decay
functions.

\begin{figure}
\begin{center}
\includegraphics[angle=0,width=0.75\textwidth]{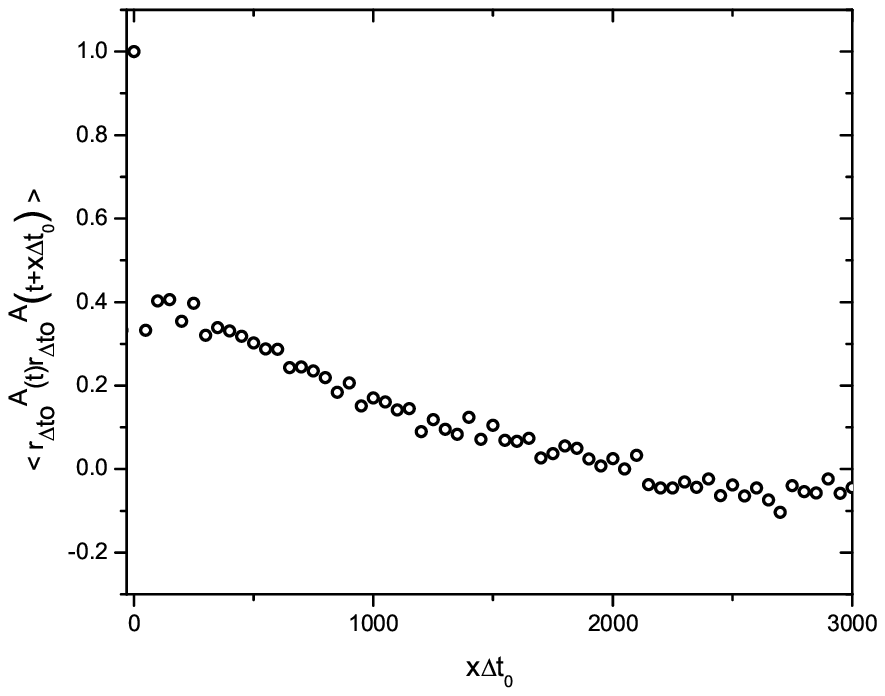}
\caption{\label{fig:autodecay} (Color online) The decay function,
$\left\langle r_{\Delta t_{0}}^{A}(t)r_{\Delta t_{0}}^{A}(t+x\Delta
t_{0})\right\rangle$ in case of the persistent random walk, sampled
with Weibull sampling intervals. The autocorrelations decrease
slowly.}
\end{center}
\end{figure}

\begin{figure}
\begin{center}
\includegraphics[angle=0,width=0.75\textwidth]{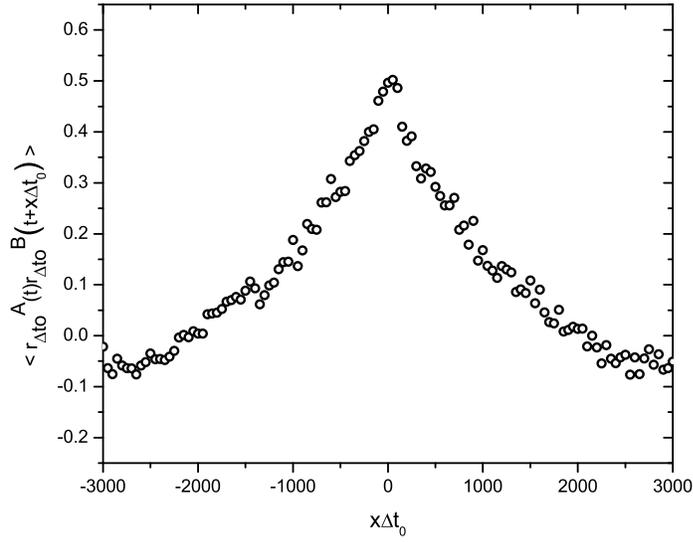}
\caption{\label{fig:cross_decay} (Color online) The decay function,
$\left\langle r_{\Delta t_{0}}^{A}(t)r_{\Delta t_{0}}^{B}(t+x\Delta
t_{0})\right\rangle$ in case of the persistent random walk, sampled
with Weibull sampling intervals. The cross-correlations decrease
slowly.}
\end{center}
\end{figure}

In the above examples we had random walks as underlying processes,
thus by definition the autocorrelations are delta functions. However,
because of the asynchronous sampling, the cross-correlation is being
smeared out and instead of a delta function we have finite decay of
the cross-correlation. These cases are simple from the point of view
of the decay functions. To demonstrate that in case of more
complicated vanishing decay functions the decomposition can still give
a good estimation of the asymptotic correlation, we consider the
following time series. We generate a persistent random walk
\cite{furth1917,weiss1994}, ie. a walk, where the probability,
$\alpha$, of jumping in the same direction as in the previous step is
higher than $0.5$. Then, as we have done before, we sample the
persistent random walk twice independently with sampling intervals drawn
from a Weibull distribution (again with parameters $a=20$ and
$b=0.7$). This construction generates slowly vanishing decay functions
(that would be exponentially decaying without the asynchronous
sampling). Again we generate 50 pairs of time series, each being 25000
steps long, the persistency is $\alpha=0.999$. Figures
\ref{fig:autodecay} and \ref{fig:cross_decay} shows two of the decay
functions (the decay of the autocorrelations are identical so we only
show one of them). Here we set $\Delta t_{0}=50$.

\begin{figure}
\begin{center}
\includegraphics[angle=0,width=0.75\textwidth]{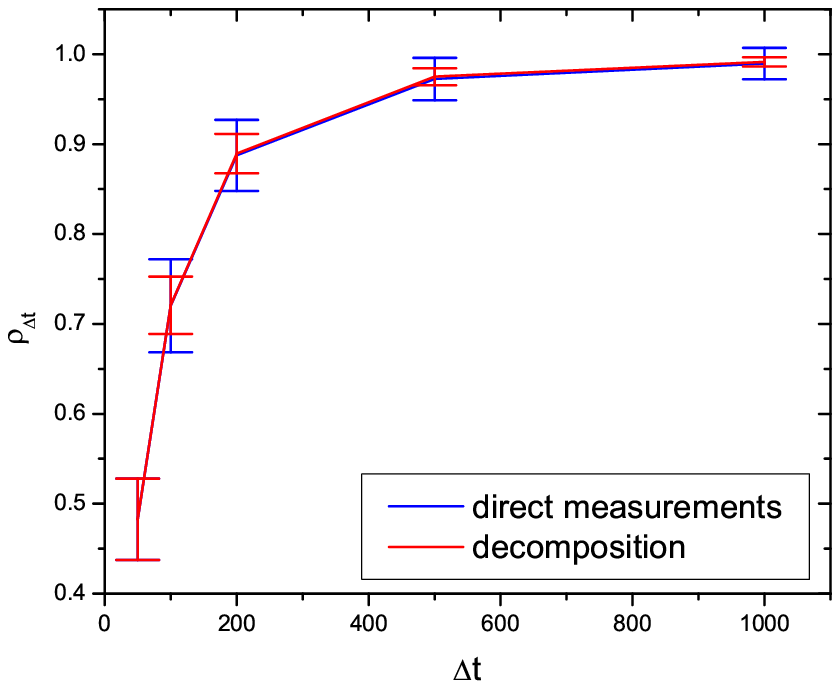}
\caption{\label{fig:weib_persistent} (Color online) Comparison of the directly measured
correlation coefficients and the coefficients determined through the
decomposition method in case of the persistent random
walk, with Weibull sampling.
In blue we show the average and the standard deviation
of the direct measurements computed over the ensemble of 50 time series
pairs. In red we show the average and the standard deviation
obtained by the decomposition method, computed over the ensemble of 50 time series pairs.}
\end{center}
\end{figure}

\begin{figure}
\begin{center}
\includegraphics[angle=0,width=0.75\textwidth]{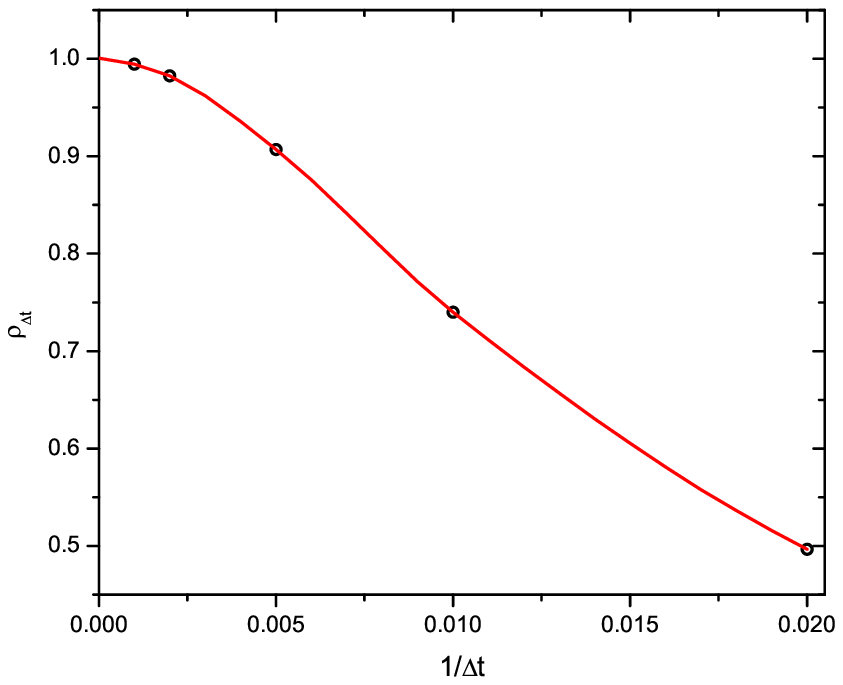}
\caption{\label{fig:weib_pers_pchip} (Color online) The correlation
as a function of $1/\Delta t$ for the persistent random walk, with
Weibull sampling. The circles show the correlations determined using
the decomposition method, and the (red) curve is determined by using
piecewise Cubic Hermite Interpolation method. The extrapolation gives
an asymptotic correlation value of 1.002.}
\end{center}
\end{figure}

In Figure \ref{fig:weib_persistent} we show the results for the
correlation coefficients on different time scales. The decomposition
method gives good results in this case too. The ratio of the standard deviations
at $\Delta t=1000$ is close to 3.5, signaling that in order to obtain the same precision, we need
roughly one order of magnitude more data points in case of direct measurements than
for the decomposition method.
Figure \ref{fig:weib_pers_pchip} shows the extrapolation to the
asymptotic value of the correlation, using piecewise Cubic Hermite
Interpolation method.  The extrapolated curve intercepts the y-axis at
the value of 1.002.
Applying the extrapolation to the endpoints of the error bars, comparing the
direct measurements and the decomposition results, we find a factor of
20 improvement in the precision.

\section{\label{sec:data} Demonstration on real data}
As discussed in the introduction, an example of correlated asynchronous signals
is the case of stock market data. Price changes for different assets on
the market appear in an asynchronous manner, however, it is well known that there
are important correlations between the price changes. In order to demonstrate
our method on real world data, in this Section we show how it can be used
to estimate financial correlations.

We took data for Coca-Cola and Pepsi, a pair of stocks with strong correlations from the high frequency Trade and Quote (TAQ) Database of the New
York Stock Exchange (NYSE) for the year 2000. We computed the logarithmic
returns of stock prices:

\begin{equation}
\label{eq:fin_ret}
q^{A}_{\Delta t}(t)=\log\frac{p^{A}(t)}{p^{A}(t-\Delta t)},
\end{equation}

\noindent where $p^{A}(t)$ stands for the price of stock $A$ at time $t$.
The prices were determined using previous tick estimator on the high frequency
data, i.e. prices are defined constant between two consecutive trades. What we
study is the cross-correlation coefficient between the data of different stocks.

We divided data for the year 2000 into 50 disjoint
periods of 5 days (weekly periods) and measured the correlations on these time
intervals. This way we handle the separate weeks independently, similar to the
case of the generated data: We have $50$ time series pairs and can study
both the directly calculated correlation coefficients, both the coefficients
obtained through the decomposition method.

\begin{figure}
\begin{center}
\includegraphics[angle=0,width=0.75\textwidth]{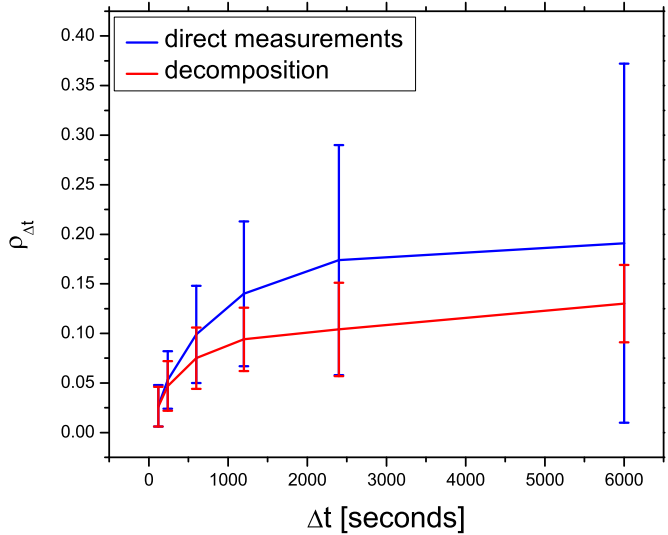}
\caption{\label{fig:data} (Color online) Comparison of the directly measured
correlation coefficients and the coefficients determined through the
decomposition method in case of real world stock market data.
The correlations are computed for the stock pair KO/PEP for the year 2000.
In blue we show the average and the standard deviation
of the direct measurements computed over the ensemble of 50 weekly time series
pairs. In red we show the average and the standard deviation
obtained by the decomposition method, computed over the ensemble of 50 weekly time series pairs.}
\end{center}
\end{figure}

Figure \ref{fig:data} shows the results for the correlation coefficients
on different time scales.
We can see that the standard deviation of
the coefficients obtained through the decomposition are much lower than
those of the direct measurements: The ratio of the standard deviations
at $\Delta t=6000$ seconds is 4.6 signaling that in order to obtain the same precision, we need
roughly 20 times more data points in case of direct measurements than
for the decomposition method.

\section{\label{sec:summary} Summary}
In this paper we discussed the problem of estimating the correlation
coefficient between two asynchronous signals. While the direct use of
high resolution data results in an underestimation of the
correlations, coarser binning of the the data leads to larger errors
due to loss of data.

We proposed a method, which enables to estimate the asymptotic value
of correlations from the high frequency data, without the need of
using longer time scales and thus without using worse statistics. The
correlations from the high frequency data can be determined very
accurately, based on the good statistics.  We demonstrated our method
on generated data sets, showing that the error of correlations
determined by our method is much smaller than the errors of
correlations measured directly, using long time windows. Extrapolating
to the asymptotic correlation from the determined correlation values
leads to a very accurate estimation of the underlying correlation. A
very important question in the estimation of the asymptotic
correlation value is the determination of the shortest meaningful time
scale, $\Delta t_{0}$, on which we measure the decay functions. The
asynchronicity of the signals slows down the decrease of the decay
functions. In the paper we showed that also in case of non-trivial
decay functions the decomposition gives a good estimation of the
asymptotic correlation value.

We demonstrated how the method works for real data. When studying weekly
cross-correlations of stock returns we showed that the precision
of the coefficients obtained through our method is much higher than
that of the direct measurements.

\textbf{Acknowledgements}

We thank Zolt\'an Szatm\'ary and Zsolt Kajcsos for useful
discussions. Support by OTKA K60456 and T049238 is acknowledged.

\appendix*
\section{\label{sec:appendix}The decomposition of correlations}
We can write the correlation coefficient from Equation \ref{eq:C} in the following form:

\begin{eqnarray}\label{eq:app_1}
\rho_{\Delta t}^{A/B}=\frac{\left\langle \Delta r_{\Delta t}^{A}(t)\Delta r_{\Delta
t}^{B}(t)\right\rangle}{\sigma^{A}\sigma^{B}}=\nonumber\\
=\frac{\left\langle r_{\Delta t}^{A}(t)r_{\Delta
t}^{B}(t)\right\rangle - \left\langle r_{\Delta t}^{A}(t)\right\rangle
\left\langle r_{\Delta
t}^{B}(t)\right\rangle}{\sqrt{\left\langle r_{\Delta
t}^A(t)^{2}\right\rangle - \left\langle r_{\Delta
t}^A(t)\right\rangle^2}\sqrt{\left\langle r_{\Delta
t}^B(t)^{2}\right\rangle - \left\langle r_{\Delta
t}^B(t)\right\rangle^2}}.
\end{eqnarray}

We assume two time scales: $\Delta t$ and $\Delta t_0$ where $\Delta t=n\Delta t_0$, with $n$ being a
positive integer.
The change in the measured quantity in the time window $\Delta t$ is the mere sum of changes
in shorter, non-overlapping time windows $\Delta t_{0}$:

\begin{eqnarray}
\label{eq:app_ret_scale}
r_{\Delta t}(t)=\sum_{s=1}^{n}r_{\Delta
t_{0}}(t-\Delta t+s\Delta t_{0}).
\end{eqnarray}

Using this relationship the time average can be written in the following form:

\begin{eqnarray}
\label{eq:app_ave_scale}
\left\langle r_{\Delta t}^{A}(t)r_{\Delta
    t}^{B}(t)\right\rangle=\frac{1}{T-\Delta t}\sum_{i=\Delta
    t}^{T}r_{\Delta t}^{A}(i)r_{\Delta t}^{B}(i)= \nonumber \\
    =\sum_{s=1}^{n} \sum_{q=1}^{n} \left\langle r_{\Delta t_{0}}^{A}(t-\Delta t+s\Delta
    t_{0})r_{\Delta t_{0}}^{B}(t-\Delta t+q\Delta t_{0})\right\rangle.
\end{eqnarray}

The sum in Eq. \ref{eq:app_ave_scale} can be written in the following way for stationary signals:

\begin{eqnarray}
\label{eq:app_ave_assumption}
\left\langle r_{\Delta t}^{A}(t)r_{\Delta
    t}^{B}(t)\right\rangle=\sum_{x=-n+1}^{n-1}\left(n-|x|\right)\left\langle r_{\Delta
    t_0}^{A}(t)r_{\Delta t_0}^{B}(t+x\Delta t_0)\right\rangle,
\end{eqnarray}
and similarly
\begin{eqnarray}
\label{eq:app_ave_assumption_2}
\left\langle r_{\Delta t}^{A}(t)^2\right\rangle=\sum_{x=-n+1}^{n-1}\left(n-|x|\right)\left\langle r_{\Delta t_0}^{A}(t)r_{\Delta
    t_0}^{A}(t+x\Delta t_0)\right\rangle \nonumber \\
\left\langle r_{\Delta t}^{B}(t)^2\right\rangle=\sum_{x=-n+1}^{n-1}\left(n-|x|\right)\left\langle r_{\Delta t_0}^{B}(t)r_{\Delta
    t_0}^{B}(t+x\Delta t_0)\right\rangle.
\end{eqnarray}

On the other hand, for stationary signals, Equation \ref{eq:app_ret_scale} leads to

\begin{eqnarray}
\label{eq:app_ave_assumption3}
\left\langle r_{\Delta t}(t)\right\rangle=n\left\langle r_{\Delta
    t_0}(t)\right\rangle.
\end{eqnarray}

Combining the above equations we can deduce a relationship between the correlation coefficients
measured on two different sampling time scales and we get Equation \ref{eq:data_formula2}.


\end{document}